\documentstyle [12pt,aasms4] {article}
\input epsf

\begin{document}

\title{Clustering of the Diffuse Infrared Light from the COBE DIRBE
maps. An all-sky survey of $C(0)$.}

\author{A. Kashlinsky\altaffilmark{1}, J. C. Mather\altaffilmark{2}, S.
Odenwald\altaffilmark{3}}
\affil{$^1$NORDITA, Blegdamsvej 17, DK-2100 Copenhagen, Denmark}
\affil{$^2$Code 685, NASA Goddard Space Flight Center, Greenbelt, MD 20771}
\affil{$^3$Hughes STX Corporation, Code 685.3,\\ NASA Goddard Space Flight
Center, Greenbelt, MD 20771}

\def\plotone#1{\centering \leavevmode
\epsfxsize=\columnwidth \epsfbox{#1}}

\def\wisk#1{\ifmmode{#1}\else{$#1$}\fi}

\def\wm2sr {Wm$^{-2}$sr$^{-1}$ }		
\def\w2m4sr2 {W$^2$m$^{-4}$sr$^{-2}$ }		
\def\nwm2sr {nWm$^{-2}$sr$^{-1}$ }		
\def\lt     {\wisk{<}}
\def\gt     {\wisk{>}}
\def\le     {\wisk{_<\atop^=}}
\def\ge     {\wisk{_>\atop^=}}
\def\lsim   {\wisk{_<\atop^{\sim}}}
\def\gsim   {\wisk{_>\atop^{\sim}}}
\def\kms    {\wisk{{\rm ~km~s^{-1}}}}
\def\Lsun   {\wisk{{\rm L_\odot}}}
\def\Msun   {\wisk{{\rm M_\odot}}}
\def\um     {$\mu$m}
\def\sig    {\wisk{\sigma}}
\def\etal   {{\sl et~al.\ }}
\def\eg	    {{\it e.g.\ }}
\def\ie     {{\it i.e.\ }}
\def\bsl    {\wisk{\backslash}}
\def\by     {\wisk{\times}}

\def\amin   {\wisk{^\prime\ }}
\def\asec   {\wisk{^{\prime\prime}\ }}
\def\cc     {\wisk{{\rm cm^{-3}\ }}}
\def\deg    {\wisk{^\circ}}
\def\ddeg   {\wisk{{\rlap.}^\circ}}
\def\damin  {\wisk{{\rlap.}^\prime}}
\def\dasec  {\wisk{{\rlap.}^{\prime\prime}}}
\def\approxeq{$\sim \over =$}
\def\abouteq{$\sim \over -$}
\def\percm{cm$^{-1}$}
\def\percmsq{cm$^{-2}$}
\def\percmcub{cm$^{-3}$}
\def\perhz{Hz$^{-1}$}
\def\perpc{$\rm pc^{-1}$}
\def\persec{s$^{-1}$}
\def\peryr{yr$^{-1}$}
\def\te{$\rm T_e$}
\def\tenup#1{10$^{#1}$}
\def\to{\wisk{\rightarrow}}
\def\thin{\thinspace}
\def\uk{$\rm \mu K$}
\def\p{\vskip 13pt}

\begin{abstract}
We measure the smoothness of the infrared sky using the COBE DIRBE
maps, and obtain interesting limits on the production of the diffuse
cosmic infrared background (CIB) light by matter clustered like galaxies.
The predicted fluctuations of the CIB with the DIRBE beam size of 0.7\deg\
are of the order of 10\%, and the maps are smooth at the level of
$\delta\nu I_\nu \sim $ a few \nwm2sr rms from 2.2 to 100 \um.
The lowest numbers are achieved at mid- to far-IR where the foreground
is bright but smooth; they are $\sqrt{C(0)}$$\leq$$(1-1.5)$ \nwm2sr at
$\lambda=$ 10-100 \um.  If the
CIB comes from clustered matter evolving according to typical
scenarios, then the smoothness of the maps implies CIB levels less than
$\sim (10-15) $ \nwm2sr over this wavelength range.
\end{abstract}

\newpage
\section{Introduction}
The cosmic infrared background (CIB) contains information about the conditions
in the post-recombination Universe and links the microwave background,
which probes the last scattering surface, and the optical part of the
cosmic spectrum, which probes the conditions
in the Universe today at $z\sim 0$. Numerous models have been developed
to predict the properties of the CIB at wavelengths from the near-IR
($\lambda \sim $1-10 $\mu$m) to the far-IR (e.g. Bond,
Carr and Hogan 1986; Beichman and Helou 1991; Franceschini \etal 1991).
The predicted spectral properties and the amplitude of the CIB depend on the
various
cosmological assumptions used, such as the cosmological density parameter,
the history of
star formation, and the power spectrum of the primordial density field among
others. A typical prediction over the range  of wavelengths probed by
the COBE (Cosmic Background Explorer) DIRBE (Diffuse Infrared Background
Experiment),
1.25 - 240 $\mu$m, is $\nu I_\nu \sim 10$ \nwm2sr (e.g. Pagel 1993
and references therein).  It is difficult to measure such levels
directly because the foreground emissions from stars and interstellar
and interplanetary dust are bright.

An alternative to direct photometric measurements of the uniform (DC)
component
of the CIB is to study its spatial structure. In
such a method, the groundwork for
which has been laid in Kashlinsky, Mather, Odenwald and Hauser (1996;
hereafter Paper I), one compares the predicted fluctuations of the CIB
with the measured angular correlation function of the maps. Here  we apply the
method to the all-sky DIRBE maps and show that
it imposes interesting limits on the CIB from clustered matter,
particularly in the mid- to far-IR.
We review the theoretical basis for the analysis and improve
it over Paper I, discuss the data sets and the methods of map
construction and analysis, and conclude with the results and limits they set
on the CIB.

\section{Theory}
The angular correlation function for the CIB has been studied for a variety of
cosmological models (e.g. Bond, Carr and Hogan 1986, Wang 1991, Coles, Treyer
and Silk 1990, Paper I). In general, the
intrinsic correlation function of the diffuse
background, $C(\theta) \equiv \langle \nu \delta I_\nu({\bf x}) \cdot
\nu \delta I_\nu ({\bf x}+\theta)\rangle$, where $\delta I \equiv I-
\langle I \rangle$ is the map of spatial fluctuations produced by a population
of emitters (e.g. galaxies) clustered with a 3-dimensional correlation
function $\xi(r)$, is given
in the small angle limit ($\theta \ll 1$) by
\begin{equation}
C(\theta) = \int_0^\infty A_\theta(z) (\frac{d\nu I_\nu}{dz})^2
[\Psi^2(z) (1+z)^2 \sqrt{1+\Omega z}] dz,
\label{e1}
\end{equation}
where it was assumed that the cosmological constant is zero, $\Psi(z)$ is the
factor accounting for the evolution of the clustering pattern, and
$A_\theta (z)= 2R_H^{-1} \int_0^\infty
\xi(\sqrt{v^2 + \frac{x^2(z)\theta^2}{(1+z)^2}}\;) dv$. Here $R_H=cH_0^{-1}$,
and $x(z)$ is the comoving distance.  In case of non-zero cosmological
constant,
$\Lambda \equiv 3 H_0^2 \lambda$, the $\sqrt{1+\Omega z}$ term in the
parentheses above should be replaced with $\sqrt{1+\Omega z +\lambda
[(1+z)^{-2}-1]}$.
After convolving Eq. \ref{e1}
with the beam, the zero-lag correlation function (mean square deviation)
becomes
\begin{equation}
C_\vartheta(0) = \int_0^\infty A_\vartheta(z) (\frac{d\nu I_\nu}{dz})^2
[\Psi^2 (1+z)^2 \sqrt{1+\Omega z}] dz,
\label{e2}
\end{equation}
\begin{equation}
A_\vartheta(z)=\frac{1}{2\pi R_H} \int_0^\infty P_{3,0}(k) k
W(\frac{kx(z)\vartheta}{1+z}) dk.
\label{e3}
\end{equation}
For the top-hat beam of DIRBE the window function is $W(x)=[2J_1(x)/x]^2$ and
$\vartheta$=0.46\deg. In Eq. \ref{e3}, $P_{3,0}(k)$ is the spectrum of
galaxy clustering at the present epoch. If $\xi$  is known, the measurement
of $C(\theta)$ can give information on the diffuse background due to material
clustered like galaxies (e.g. Gunn 1965, Peebles 1980). Such a method was
successfully applied in the V (Shectman 1973, 1974)
and UV bands (Martin and Bowyer 1989). In the rest of the paper we omit the
subscript $\vartheta$ in Eq. \ref{e3},
with $C(0)$ referring to the DIRBE convolved
zero-lag correlation signal; the mean square of the map
$\langle (\nu \delta I_\nu)^2\rangle$.  This $C(0)$ is just the
mean square confusion noise.

As was shown in Paper I, for scales dominating the integral in Eq.
\ref{e2} the
2-point correlation function at the present epoch can be approximated as a
power-law. In that case, in the Friedman-Robertson-Walker Universe,
$A_\vartheta(z)$ would reach a minimum at $z\geq 1$ whose value is almost
independent of $\Omega$. Here we generalize the argument
to a more realistic $\xi(r)$. We computed $A_\vartheta(z)$ using the
correlation data from the APM (Maddox \etal 1990) survey, with the APM power
spectrum taken from: 1) the inversion technique of Baugh and Efstathiou (1993;
hereafter BE), and; 2) the empirical fit to the APM data on the projected
galaxy correlation function from Kashlinsky (1992; hereafter
K92). The resultant $A_\vartheta(z)$ is shown in Fig. 1 for $\Omega$=1 (upper
sets of lines) and 0.1 (lower sets). The solid lines correspond to BE and the
dashed lines to K92 spectra $P_{3,0}(k)$. Both were normalized to the
Harrison-Zeldovich spectrum at wavenumbers smaller than those probed by
the APM data, as required by the COBE/DMR observations (Bennett \etal
1996). The three
sets of curves for each $\Omega$ indicate the $ 1\sigma$ uncertainty from the
APM data from the BE inversion and a roughly similar uncertainty from the K92
empirical fit. The minimum of $A_\vartheta(z)$ is clear, and
for the scales probed by the DIRBE beam the value at the minimum is
practically independent of $\Omega$. Following Paper I we can rewrite
Eq. \ref{e2}
as an inequality to derive an upper limit on a measure of the CIB flux
from clustered material from any upper limit on $C(0)$ derived from the DIRBE
data:
\begin{equation}
(\nu I_\nu)_{z,rms} \leq B \sqrt{C(0)}
\label{e4}
\end{equation}
where $B \equiv 1/\sqrt{{\rm min}\{A_\vartheta(z)\}}= (11-14)$ over the
entire range of parameters, and the measure of the CIB flux used in Eq.
\ref{e4} is
defined as $[(\nu I_\nu)_{z,rms}]^2 \equiv \int
(\frac{d\nu I_\nu}{dz})^2 [\Psi^2(z)(1+z)^2\sqrt{1+\Omega z}] dz$.
The latter is $\simeq \int (\frac{d\nu I_\nu}{dz})^2 dz$ since the
term in the brackets has little variation with $z$ for two extremes of
clustering evolution, when it is stable in either proper or comoving
coordinates (Peebles 1980). Thus the CIB  produced by objects clustered like
galaxies should have significant fluctuations, $\sim 10\%$ of the total
flux, on the angular scale subtended by the DIRBE beam. As Eq.
\ref{e2}
shows, if the bulk of the CIB comes from higher redshifts this would lead to
smaller relative $\delta I/I$ and vice versa (cf. Wang 1991).

\section{Data and analysis}
The COBE was
launched in 1989  (Boggess \etal 1992).  The DIRBE is one of the three
instruments, and mapped the sky  in ten bands
from the near-IR (J, K, L bands at $\lambda$=1.25, 2.2 and 3.5 $\mu$m) to the
far-IR at 240 $\mu$m. Four of the bands
match the IRAS bands. The instrument is a photometer with
an instantaneous field of view of 0.7\deg$\times$0.7\deg . It was designed to
have a mission-averaged, instrumental noise less than 1 \nwm2sr in the first
eight bands; at
140 and 240 \um\ the instrumental noise is higher.

We analyzed maps for all 10 DIRBE bands derived from the entire
41 week DIRBE data set available from the NSSDC. A parametrized model
developed by the DIRBE team (Reach et al. 1996) was used to remove the
time-varying zodiacal component from each weekly map (Paper I).
The maps were pixelized using the quadrilateralized spherical cube as
described by Chan and O'Neill (1974) and Chan and Laubscher (1976).

After the maps were constructed, the sky was divided into 384 patches of
32$\times$32 pixels $\simeq 10\deg\times$10\deg\  each. Each field was cleaned
of bright sources by the program developed by the DIRBE team, in which the
large scale flux
distribution in each patch was modeled with a fourth order polynomial.
As in Paper I, pixels with fluxes $> N_{\rm cut}$ standard
deviations above the fitted model were removed along with the surrounding 8
pixels. Three values of $N_{\rm cut}$ were used: $N_{\rm cut}$ =7, 5, 3.5.
Since
any large-scale gradients in the emission are clearly due to the local
foregrounds, a fourth order polynomial was also removed from each patch after
desourcing. The extragalactic contribution to $C(0)$ should come predominantly
from small scales, so removing large scale gradients would not remove any
significant part of the extragalactic $C(0)$. In any case,
we confirmed that removing the local gradients with lower order
polynomials makes little difference in the final results.
We also verified that there is a good
correlation between  removed bright objects in the 1.25, 2.2, and 3.5
\um\ bands
at $N_{\rm cut}=7 $ and the stars
from the SAO catalog; at lower $N_{\rm cut}$ many of the removed peaks
would be too dim to enter the catalog.

The results of the calculations of $C(0)$ for all 384 patches at
$N_{\rm cut}=3.5$ for wavelengths from 1.25 to 100 \um\
are shown in Plate 1 in Galactic
coordinates. Sixteen shades are used with a logarithmic increment of
$\simeq$1.8 in $C(0)$ starting at the minimal values in each band.
The fluctuations in all bands are strongest where the
foregrounds are brightest in the Galactic or Ecliptic plane.
The histograms for the $C(0)$ maps in each band show that the
range of values for $C(0)$ are very different
in the near-IR (1.25 - 4.9 \um) and mid- to far-IR (12-100 \um)
bands.  In the near-IR where Galactic stars dominate the fluctuations
of the foreground, there is the expected dependence of the minimal value
of $C(0)$  on the desourcing parameter $N_{\rm cut}$.
  At mid- to far-IR there is very little change
in the distribution of $C(0)$ with $N_{\rm cut}$ since
the foreground emission is extended with little small-scale structure.

In the near-IR bands where Galactic stars dominate the foreground there
is a strong correlation between the residual value of $C(0)$ and the
Galactic latitude: $C(0)$$\propto$$(\sin |b|)^{-\alpha}$ with
$\alpha$$\simeq$2. From least square fits we found $\alpha$=2.2, 2.4, 2.5 and
1.7
for 1.25, 2.2, 3.5, and 4.9 \um\ respectively.
At longer wavelengths the correlation with $b$ is significantly less
pronounced due to appreciable contributions from the zodiacal
foreground.  At short
wavelengths (1.25 - 4.9 \um) the remaining fluctuations are still
mainly due to point sources, which are recognizable as small groups of
bright pixels in the original maps.  This is confirmed by inspection of
the histograms of the pixel brightnesses, which are asymmetric as
expected for a distribution of point sources at various distances (Paper
I, Fig. 5).  Because point sources clearly dominate the fluctuations,
the choice of $N_{\rm cut}$ is important in the near-IR.  On the other hand,
removal of large scale gradients using a variety of polynomial fits after
removing point sources makes little difference to the calculation of
$C(0)$ for these bands.

At longer wavelengths the situation is quite different.  The sky
brightness is dominated in most directions outside the Galactic Plane
by the interplanetary dust, which is very smoothly distributed except
for a cusp in the Ecliptic plane and some faint dust bands within a few
degrees of it.  There are also clouds of dust grains in resonance with
the Earth's orbit, which appear in the Ecliptic plane 90\deg\ from the
Sun (Reach \etal 1995b).  Outside these regions, the main foreground structures
are interstellar dust, which has typical size scales larger than the
DIRBE beam ( Waller and Boulanger, 1994; Low and Cutri, 1994).
Therefore it makes a difference whether
large scale gradients are removed before computation of $C(0)$. The
distribution of pixel brightnesses is also more nearly Gaussian and
much less asymmetrical than at short wavelengths, and the choice of the
desourcing parameter $N_{\rm cut}$ makes little difference.

Because the measured fluctuations in the mid-IR are so small relative to
the total brightness, and point sources are not the obvious source of
fluctuations, we must evaluate  instrument
noise effects.  Inspection of the raw data shows that the detector noise is
comparable to the digitization roundoff noise for each
detector sample except in Bands 9 and 10 where the detector noise
dominates.  The onboard phase sensitive detection algorithm
obtains 8 digital samples per chopper cycle
of the analog preamplifier outputs. The DIRBE beam scan rate is such
that four of these chopper cycles
are averaged to obtain a single photometric measure of the sky
brightness at each each pixel location.
The rms conversion noise is $12^{-0.5}$ digital units per sample, so that
for a single pixel measurement consisting of  32 digital values, one
obtains an rms digital noise of 0.051 digital units. The preamplifier
gain is
16 for the sky observations, so that the conversion of the digital units
to photometric units yields  digitization noise levels of 3.2, 1.7, and
0.4  \nwm2sr in the J, K and L bands, 0.6 - 0.9 \nwm2sr for the bands between
4.9 and 100 \um\  , and 1.7-5.6 \nwm2sr in the 140 and 24 \um\ bands. These
digitization noise estimates are further reduced by a factor of 17 by averaging
the individual photometric measures for each pixel, since
each pixel is observed about 15 times per week, and 20 weeks per year.
Regions near the
ecliptic poles are observed substantially more frequently.
   During this time the
brightness of the interplanetary dust changes dramatically because of
the Earth's motion through the dust cloud, so wide ranges of the
digitizer are exercised.  Therefore the digitization errors are
randomized so that they do not introduce obvious band structures in the
completed annual map.  We verified this from an average map made without
subtracting a model for the interplanetary dust.

On the other hand, we make a separate correction for the interplanetary
dust contribution for each weekly map.  Systematic errors in the
corrections can produce artificial stripes in the computed average map,
which are seen most easily in the Ecliptic plane where the signals
and their rates of change are greatest.  These stripes have not been
entirely eliminated by the dust model we used, and are one reason why
the $C(0)$ maps show increases near the Ecliptic plane even though the
dust emission is expected to be very smooth.  These stripes have least
effect at the Ecliptic poles.

We conclude that the measured values for the fluctuations are upper
limits on the fluctuations of the CIB.  The upper limits estimated below
come from the quietest patch in each band for the entire sky. For brevity
we omit the histogram distribution of $C(0)$, but its inspection shows that
the minimum is estimated from around 30-80 patches in each of the bands
making it a reliable and unbiased estimate of the true upper limit. Since
the distribution of $C(0)$ is highly anisotropic on the sky, the signal must
come from the foreground emission and it is thus further unlikely that
any of the truly cosmological contributions have been removed in the
process. Plate 1 shows that the lowest $C(0)$ for the adjacent bands
come from neighbouring patches. In the near IR these are located around
Galactic poles; in the mid-IR they are near Ecliptic poles and in the
far-IR they shift back to the Galactic poles. Our data processing does not
remove real fluctuations unless they appear to come from point sources,
or to have very large scale gradients.  The instrument digitization
noise, estimated to be below 1.0 \nwm2sr in all bands after averaging,
is not a problem at the level probed here.

\section{Results and conclusions}

The measured upper limits on the fluctuations of the CIB offer a
powerful test of any model for its sources.  Fig. 2 summarizes our results
at all except the two longest wavelength DIRBE bands. Triangles show the upper
limits we set on $\sqrt{C(0)}$. These upper limits were calculated by
selecting the sky patch with the lowest value of $C(0)$ from among the
ensemble of 384 patches. Translating these into limits on the uniform
part of the CIB is somewhat theory-dependent. Assuming typically
10\% fluctuations on the scale of the DIRBE beam, our results lead to
quite low upper limits on the CIB fluxes from clustered material,
particularly at mid- to far-IR bands.   They are less interesting at
140 and 240 \um\ because of higher DIRBE detector noise.
The darkest DIRBE sky limits from weekly maps are presented
with $\times$-s as reported by DIRBE (Hauser 1993). (A typical pixel
is observed for 20 weeks during the mission).
The vertical bars show the residual fluxes from the fits
reported by Hauser (1996a)
after removal of Galactic and zodiacal foregrounds, but they were not
described as detections because of uncertainties in the foreground
modeling.

At wavelengths $ >$$10$\um\ they are comparable to the levels that may
be derived from the interpretation of the spectra of $\gamma$-ray
sources, where gamma rays interact with the CIB by pair production.  One
method of calculation gives the result as $\nu I_\nu$$\simeq$$6
h (\frac{\lambda}{\mu \rm m})^{0.55}$\nwm2sr (Dwek and Slavin 1994),
plotted with solid lines for h=$H_0$/100kmsec$^{-1}$Mpc$^{-1}$=1 (upper
line) and 0.5. If these are treated as detections our upper limits on
$C(0)$ would suggest that the bulk of the CIB at these wavelengths is
produced at very high $z$, leading to the decrease of $(\nu
I_\nu)_{z,rms}$ for given $\nu I_\nu$. In other words this
would require $\delta I/I$$\leq$$2$$\times$$10^{-2} h^{-1}
(\frac{\lambda}{40\mu{\rm m}})^{-0.55}$, so the background must come
from high redshifts.  An alternative interpretation would be that the
CIB estimated from the $\gamma$-rays is not produced by clustered material.
The latter is however unlikely since the CIB is expected to be produced
by material that collapsed due to local gravity, and hence its structure
should reflect that of the primordial density field.

At the J, K, L (1.25 - 3.5 \um) bands the no-evolution (Paper I) or
reasonable evolution (Veeraraghavan 1996) models of the zero lag
correlation function in the CIB produced by normal galaxies give
numbers for $\sqrt{C(0)}$ around (1.5-3, 0.5-1.5, 0.5-1.5) \nwm2sr
respectively. Our limits at 1.25 \um\ are almost an order of magnitude
higher, but at 2.2 and 3.5 \um\ they are comparable in magnitude. If
there were an extra population of galaxies at early times they would
contribute an additional signal (see e.g. the Cole, Treyer and Silk
(1992) model designed to fit simultaneously the deep K-  and B-counts).
 Any other sources of emission that may have been active in the early
Universe in addition to normal galaxies would increase the above
theoretical estimates.

Further improvements in the analysis of the DIRBE maps are possible, as
are new instruments or new applications of old instruments.  They may
yield the first clear sign of a cosmic infrared background radiation.

\acknowledgements
We are grateful to Carlton Baugh for providing us with
the APM power spectrum data, and to Michael Hauser for a careful reading
of an early draft of this paper.
This work was supported by NASA Long Term Space
Astrophysics grant. The DIRBE software team developed the data
sets and the software used for removing sources and interplanetary dust
models.  The National Aeronautics and Space Administration/Goddard
Space Flight Center (NASA/GSFC) is responsible for the design,
development, and operation of the Cosmic Background Explorer (COBE).
GSFC is also responsible for the development of the analysis software
and for the production of the mission data sets.  The COBE program is
supported by the Astrophysics Division of NASA's Office of Space
Science and Applications.

\newpage

\noindent{\bf References}

\leftskip 1pc
\parindent -1pc

Baugh, C.M. and Efstathiou, G. 1993, MNRAS, {\bf 265}, 145.

Beichman, C.A. and Helou, G. 1991,  Ap.J.,  {\bf 370},  L1.

Bennett, C.L. \etal 1996, Ap.J., accepted.

Bernard, J.P. \etal 1994, A\&A, {\bf 291}, L5.

Boggess, N.W. \etal 1992, Ap.J., {\bf 397}, 420.

Bond,  J.R. \etal 1986,   Ap.J.,  {\bf 306},  428.

Bond, J.R. \etal 1991, Ap.J., {\bf 367}, 420.

COBE 1995a, COBE Skymap Information, National Space Sciences
Data Center,\\
http://www.gsfc.nasa.gov/astro/cobe/skymap\_\_info.html

COBE 1995b, DIRBE Explanatory Supplement, eds. M.G. Hauser, T.
Kelsall, D. Leisawitz, and J. Weiland, National Space Sciences Data
Center, anonymous FTP from nssdca.gsfc.nasa.gov.

Cole,  S., Treyer, M. and Silk, J.  1992,  Ap.J., {\bf 385}, 9.

Chan, F.K. and O'Neill 1975, Computer Sciences Corporation EPRF Report
2-75.

Chan, F.K. and Laubscher, R.E. 1976, Computer Sciences Corporation EPRF Report
3-76.

Dwek, E. and Slavin, J. 1994, Ap.J., {\bf 436}, 696.

Franceschini, A. \etal 1991, Ap.J.Suppl., {\bf 89}, 285.

Gunn, J. 1965, Ph.D. Thesis, Caltech. (unpublished)

Hauser, M. \etal, 1984,  Ap.J. (Letters), {\bf 278}, L15.

Hauser, M. 1993, in ``Back to the Galaxy," AIP Conf. Proc. {\bf 278},
eds. S. Holt and F. Verter, (AIP:NY), 201.

Hauser, M. 1996a, in Proc. IAU Symposium 168, ``Examining
the Big Bang and Diffuse Background Radiations", M. Kafatos and Y. Kondo, eds.,
Kluwer, Dordrecht., p. 99.

Hauser,  M. 1996b, in ``Unveiling the Cosmic Infrared
Background," AIP Conf. Proc., {\bf 348},
ed. E. Dwek, (AIP:NY), 11-24.

Kashlinsky, A. 1992,  Ap.J., {\bf 399}, L1.

Kashlinsky, A., Mather, J., Odenwald, S. and Hauser, M. 1996, Ap.J.,
accepted (Paper I)

Koo, D. and Kron, R.R. 1992, ARAA, {\bf 30}, 613.

Low, F.J and Cutri, R.M. 1994, IR Physics and Tech. 35, 291.

Maddox,  S. \etal 1990, MNRAS, {\bf 242}, 43P.

Martin, C. and Bowyer, S. 1989, Ap.J, {\bf 338}, 677.

Pagel, B.E.J. 1993, in ``The Cold Universe," eds. Montmerle \etal,
Editions Frontieres, p.345

Peebles,  P.J.E. 1980,  ``Large Scale Structure of the Universe," Princeton
Univ. Press.

Reach, W. \etal, 1995b, Nature, TBD, TBD. 

Reach, W. \etal, 1996, in ``Unveiling
the Cosmic Infrared Background," 1996, AIP Conf. Proc., {\bf 348}, (AIP:
New York), 37-46.

Veeraraghavan, S. 1996, in ``Unveiling
the Cosmic Infrared Background," AIP Conf. Proc, {\bf 348}, (AIP: New
York), 122-126.

Waller, W. and Boulanger, F. 1994, ASP Conference Series, v. 58, 129.

Wang, B. 1991, Ap.J., {\bf 374}, 465.

Weiland, J.L. \etal, 1996, in ``Unveiling
the Cosmic Infrared Background," AIP Conf. Proc. {\bf 348}, (AIP: New
York), 74-80.

White, R.A., and Mather, J.C. 1991, ``Databases from the Cosmic
Background Explorer (COBE)," Databases \& On-line Data in
Astronomy.  Astrophysics and Space Science Library, eds. M.A.
Albrecht and D. Egret, (Dordrecht: Kluwer), 171, pp. 30-34.

\newpage

{\bf Figure captions}\\
Fig. 1. $A_\vartheta(z)$ is plotted vs $z$ for 1) zero cosmological
constant with $\Omega$=1 (uppermost curves) and $\Omega$=0.1
(lowermost curves) and 2) flat Universe with $\Omega =0.1$ and
non-zero cosmological constant (middle curves). Dashed lines
correspond to the APM fit spectrum from K92 and
solid lines to BE. Three sets of each line correspond to one-sigma uncertainty
in the BE fit and approximately the same uncertainty for K92.\\
\\

Fig. 2. Triangles show the upper limits on $\sqrt{C(0)}$ set by our analysis.
$\times$-s are the lowest upper limits from the darkest part of the
DIRBE weekly sky maps from
(Hauser 1993).  Vertical bars show the range of the residual DIRBE fluxes
left after Galaxy modelling and removal (Hauser 1996a,b).
Solid lines are the possible detections from Dwek and
Slavin (1994); the upper line is for $H_0$=100 km/sec/Mpc and the lower
one is for 50km/sec/Mpc. All fluxes are in \wm2sr .\\
 \\
Plate 1. Maps of $C(0)$ in Galactic coordinates for DIRBE Bands 1-8,
1.25 - 100 \um\ . The grey scale shows logarithmic steps at intervals of 1.8.
For Bands 1-8, the minimum log($C(0)$) are -4.2, -4.7, -5.1, -5.4, -6.0,
-4.1, -3.3 and -2.5 respectively.

\clearpage
\begin{figure}
\centering
\leavevmode
\epsfxsize=1.0
\columnwidth
\epsfbox{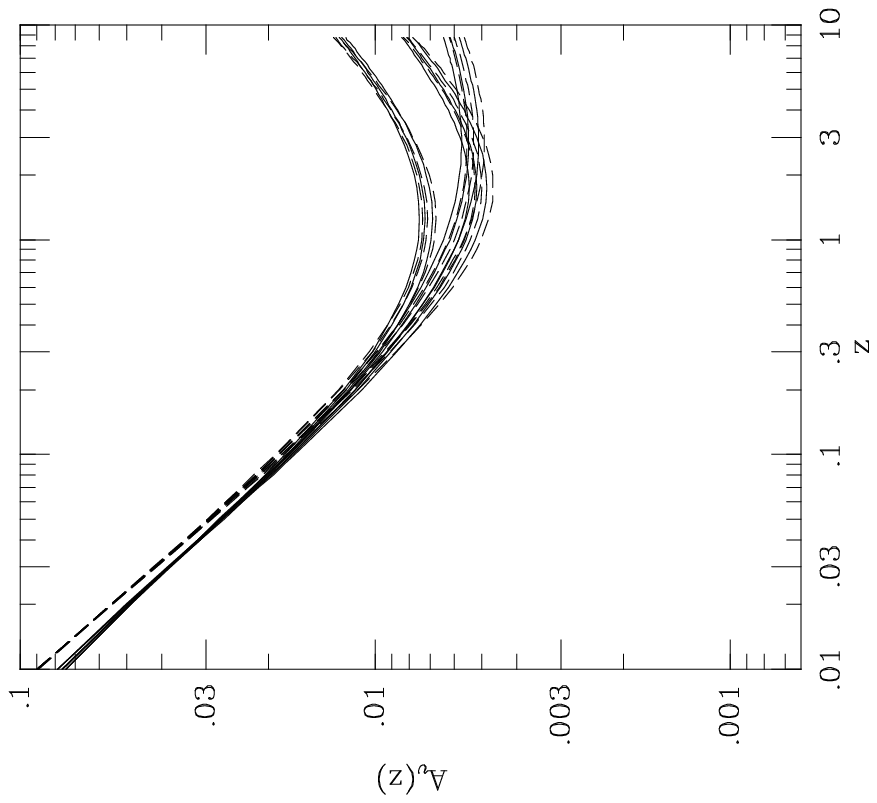}
\caption[]{}
\end{figure}
\clearpage
\begin{figure}
\centering
\leavevmode
\epsfxsize=1.0
\columnwidth
\epsfbox{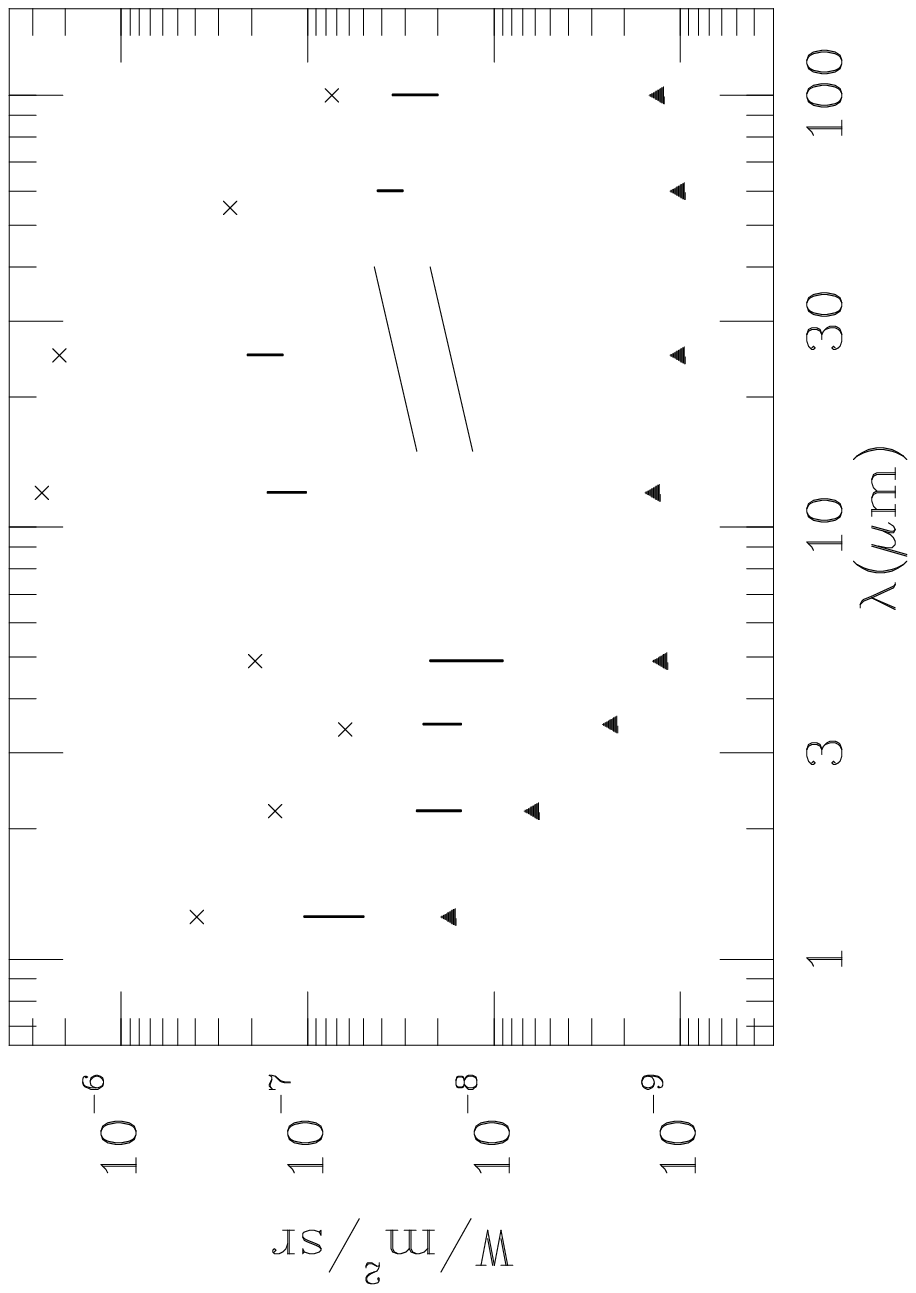}
\caption[]{}
\end{figure}
\clearpage
\begin{plate}
\centering
\leavevmode
\epsfxsize=1.0
\columnwidth
\epsfbox{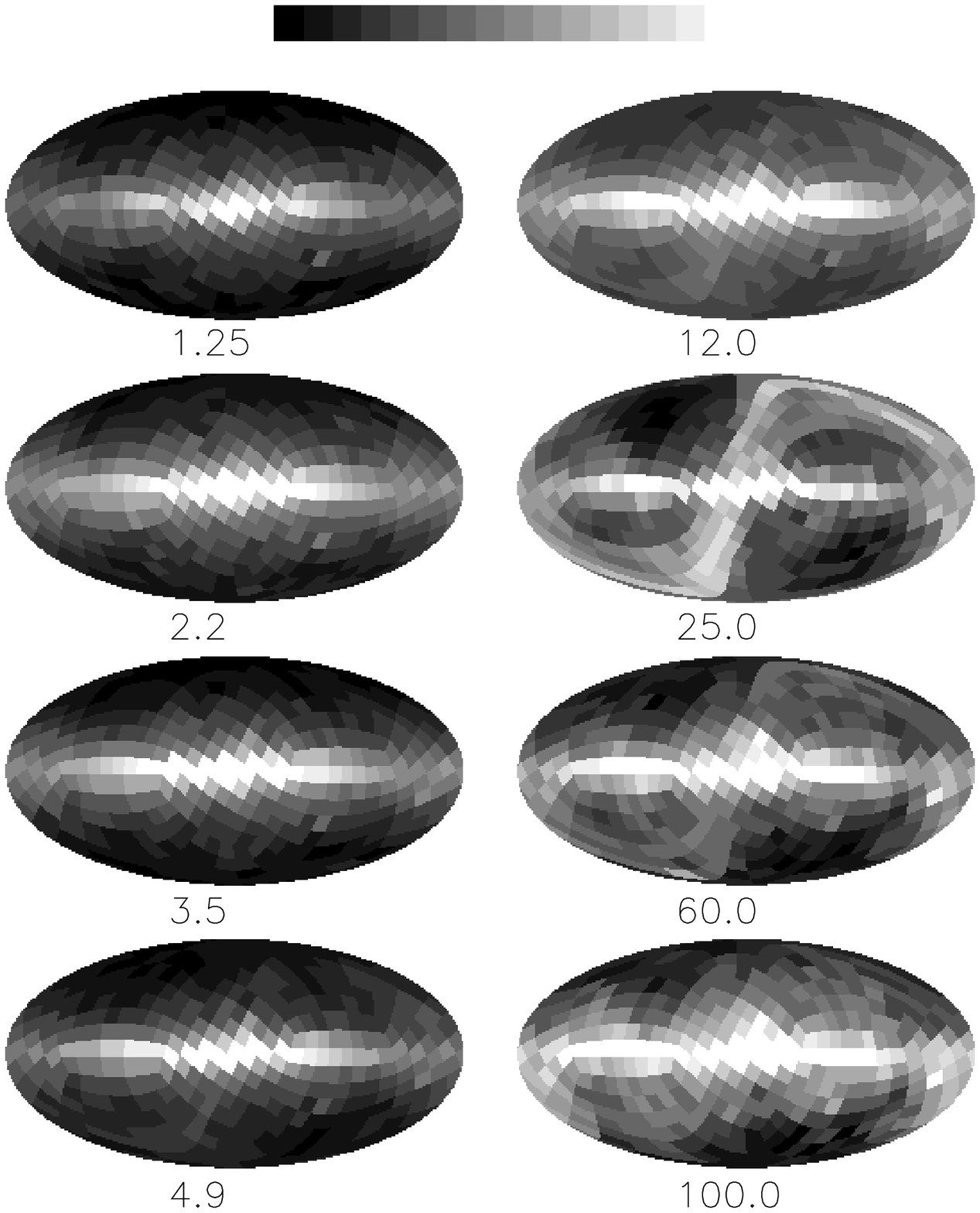}
\caption[]{}
\end{plate}
\end{document}